# THE RF-SYSTEM OF THE NEW GSI HIGH CURRENT LINAC HSI

G. Hutter, W. Gutowski, W. Hartmann, G. Kube, M. Pilz, W. Vinzenz, GSI, Darmstadt, Germany

*Abstract*

The RF part of the new high current injector-linac HSI consists of five cavities with the new operating frequency of 36 MHz instead of 27 MHz of the removed Wideroe type injector. The calculated power requirements of the cavities including beam load in three structures were between 110 kW for a rebuncher and 1.75 MW pulse-power for the two IH-cavities. The beam load is up to 150 kW for the RFQ and up to 750 kW for the two drift tube tanks. An additional 36 MHz debuncher in the transfer line to the Synchrotron (SIS) will need 120 kW pulse power. We decided to fulfil these demands with amplifiers of only two power classes, namely three amplifiers with 2 MW and six amplifiers with 200 kW pulse output power. The latter ones are also used as drivers for the 2 MW stages. The 200 kW amplifiers were specified in detail by GSI and ordered in the industry. The three 2 MW final amplifiers were designed, constructed and built by GSI. The paper gives an overview of the complete RF system and the operating performance of amplitude and phase control with beam load. It further describes some specialities of the new 2 MW amplifiers like the simplicity of the anode circuit, a very sophisticated socket for a cathode driven amplifier with cathode on dc ground, the parasitic mode-suppression, shielding and filtering of unallowable RF-radiation and operating experience since October 1999.

## 1 INTRODUCTION

Within the beam intensity upgrade program at GSI[1], the old Wideroe type injector with four tanks, working at 27 MHz, was replaced in 1999 by the high current injector HSI. During the last year of operation, one of the old 27 MHz amplifiers, originally built by Herfurth GmbH and redesigned by the Unilac-RF-group in 1984, was replaced by a 200 kW amplifier, manufactured by Hüttinger Elektronik GmbH, Freiburg[2]. In this way we could remove the old amplifier during beam-operation of the Unilac and gain a powerful plant for prototype activities of the future 2 MW amplifiers, as a connection to a 24 kV anode supply was available there.

We changed the working frequency of a second 200 kW Hüttinger amplifier from 27 MHz to 36 MHz to get a prototype driver. In parallel, five 200 kW amplifiers where built at Thomcast AG Turgi, three of them with all the electronics needed for the final 2 MW stages except the anode supply and the filament transformers[3].

A prototype of the 2 MW amplifier was brought to operation until the end of December 1998. After a redesign of the prototype three final amplifiers were built at GSI in 1999. The first of them delivered RF to the RFQ just in time, the others for the IH-structures were ready well before the tanks.

Figure 1 shows the complete 36 MHz RF distribution.

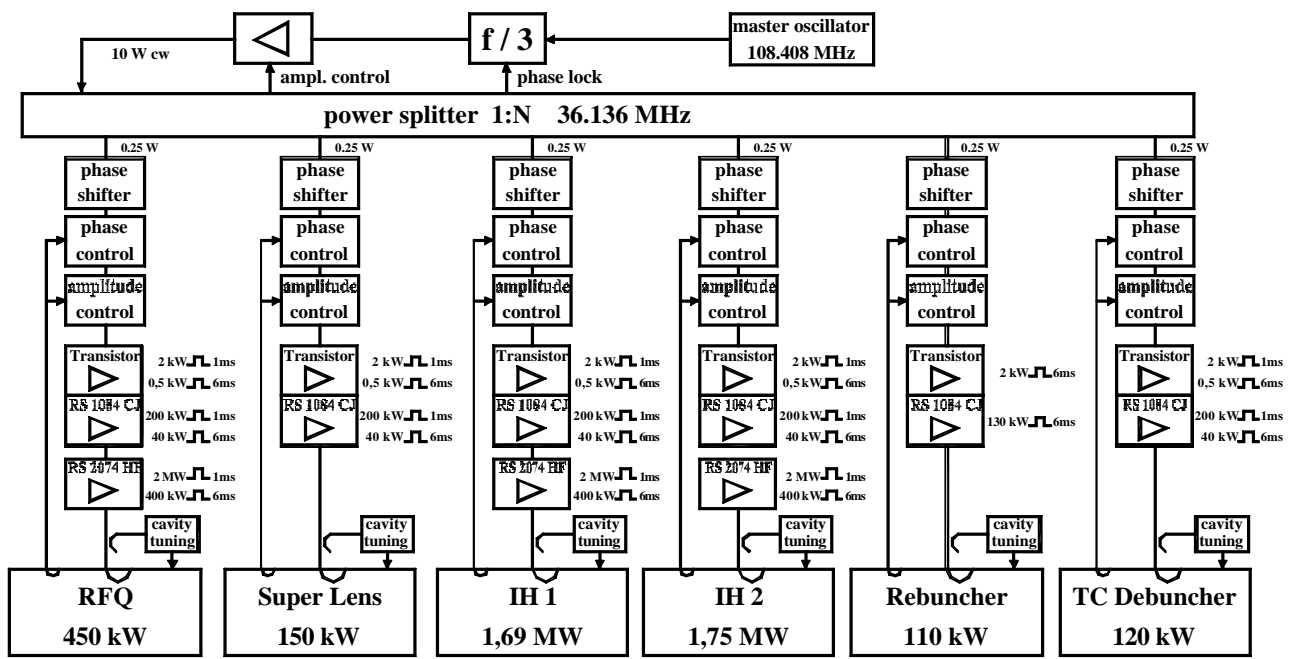

Figure 1: 36 MHz High Current Injector RF-System

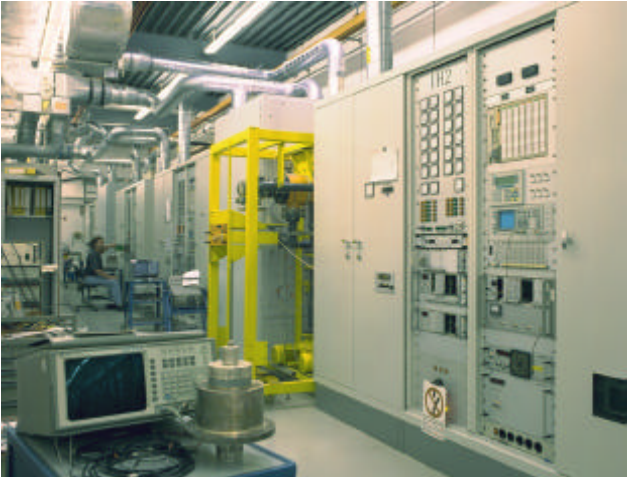
Figure 2: Overview of the new installation

## 2 THE 200 KW AMPLIFIERS

### 2.1 New amplifiers from Thomcast AG, CH

Most details of the five new 200 kW amplifiers have been reported in [3]. In 1999 all five amplifiers were installed at GSI (after the old Wideroe-amplifiers were removed) and brought to operation without remarkable problems after introducing additional damping for higher frequencies through GSI staff in the anode circuits. The only massive breakdowns occurred at the dc-power-supplies of the solid state 2 kW drivers, which had been built by a subcontractor of Thomcast. The original power-supplies have all been replaced meanwhile, but the new ones still have problems and only with the use of a spare-amplifier we could provide continuos operation.

### 2.2 Modified old amplifiers

Two 27 MHz / 200 kW amplifiers built by Hüttinger Elektronik GmbH, Freiburg, Germany[2] were designed for 10 % duty cycle. As the anode transformers of these amplifiers have 75kVA, there was a very good safety margin to use one of these amplifiers for a buncher at the end of the new injector that needs up to 30 % duty cycle, but only about 100 kW. For this purpose we had to change the frequency of the amplifier to 36 MHz. The Π-filter anode circuit could be easily brought to the proper frequency by replacing its inductance by a strip transmission line and a different tuning of the two variable vacuum capacitors. The input circuit could be tuned without changing any hardware. To suppress self-oscillations near the fundamental frequency, a neutralisation had to be introduced to the amplifier, which additionally isolated the input from the output by about 17 dB. The second amplifier was changed to 36 MHz, too, and will be used as a driver for the old 2 MW prototype and also as spare unit.

## 3 FINAL 2 MW AMPLIFIERS

### 3.1 Tube Selection

As the Siemens tube RS2074HF is installed in the five amplifiers of the Alvarez-section of the Unilac and delivers reliable 1.6 MW at 108 MHz with 25% duty cycle, we decided to take its low frequency version, the RS2074SK for the new 2 MW amplifiers. This tube is identical to the HF-version, but misses a lower-lossy-material for the screen-grid contact area, where the pyrolitic graphite grid is mounted to its support.

### 3.2 Input Circuit

From the old installations of the Wideroe-RF, the 1MVA / 24kV anode-power-supply was recuperated, as it fit all requirements with a few small changes. The reuse of the power-supply, however, implicated that the cathodes of the new amplifiers had to remain on dc-ground. On the other hand, we did not dare to build grid-driven amplifiers with the given higher frequency and power level, but chose a grid1 / grid2 based circuit with cathode drive. The RF-potential on the cathode necessitated to feed the heating-current over two parallel λ/4 strip-lines with a strong capacitive coupling against each other. On the RF-cold ends, one of them is connected to ground while the other one is connected to the heating transformer via a 1200 A filter.

Figure 3 shows in a schematic diagram, how the input-impedance of the tube is transformed to 50 Ω.

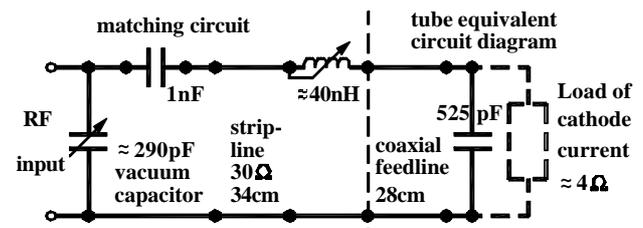
Figure 3: Equivalent input circuit diagram

Figure 4 gives a comparison of a Supercompact[Â] calculation and an analyzer measurement of the input impedance. Five parallel low inductive 20 Ω resistors simulated the cathode current in the measurement.

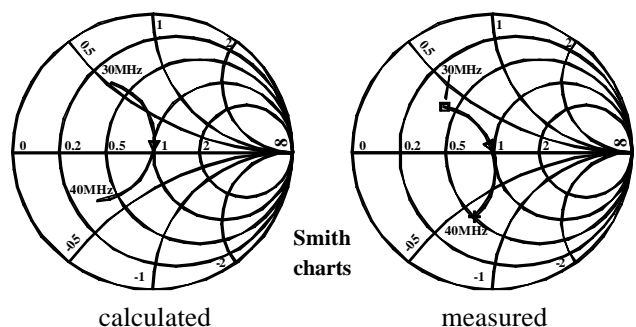
Figure 4: Input impedance from 30 MHz to 40 MHz.

### 3.3 Tube Socket

Controlgrid and screengrid are both twice capacitively blocked in the socket as close as possible to the rf-ground without any tunability. The capacitors used for this purpose were especially developed by the Swiss company Güller together with GSI. They use a copper-silver-plated polyester foil that is covered with two silver-plated brass plates, while all isolating foil parts are covered with silicone-rubber. By this treatment the capacitors are mechanically robust and absolutely waterproof. They were tested up to 8 kV successfully without partial discharges.

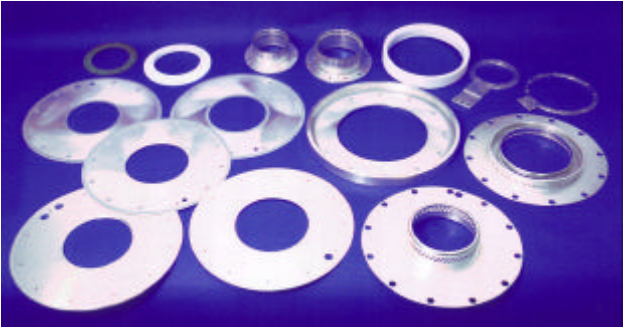

Figure 5: Parts of the sophisticated socket

When we started operation of the prototype, the output power was first limited to about 1 MW by excessive G1- and G2 current, caused by a high level of the third harmonic in the input circuit, as this frequency is near the $\lambda/4$ resonance of the G1-G2-circuit. Introducing a ring of ferrite between G1 and G2 inside the socket and changing the length of the coaxial line to the driver solved the problem and led to the data given in table 1.

### 3.4 Anode Circuit

The anode circuit is a $\lambda/4$ resonator with a shrinked in house designed and built cylindrical capton capacitor as anode-dc-blocker, which was tested up to 50kV.

Strong galvanic output coupling led to the needed low loaded-Q-value of 4, which means a 3 dB bandwidth of about 9 MHz. This makes tuning of the once adjusted anode circuit unnecessary, even after a tube change.

### 3.5 Filtering and Shielding

Unwanted RF-radiation of the new amplifiers was about 60 dB lower than the radiation of the old Unilac amplifiers, due to an excessive RF-contact-design of all circuits (there are up to 7 contacts in series) and to sufficient filtering of all dc-power-lines, including especially developed ferrite filled high voltage cables.

### 3.6 Damping of Parasitic Oscillations

When pulsing the G1 to lower values without RF-drive, the tube started parasitic oscillations around 900 MHz. As some tubes, especially an also tested RS2074HF, produced these modes even with RF, we placed 6 ferrite rods around the anode ceramic, as this had shown sufficient damping for a RS2074SK in an earlier test which GSI-staff did at HIMAC. As some RS2074HF still oscillated and as the ferrite kept cold at full power, we introduced 30 rods to each amplifier. Now both tube types are usable without oscillations.

### 3.7 Operating Parameters

Table 1: Reached Data of two operational modes

| Pulse Output Power | 400 kW | 2 MW |
|---|---|---|
| Repetition Frequency | 50 Hz | 20 Hz |
| Duty-Cycle | 30 % | 6.0 % |
| Average Output Power | 120 kW | 120 kW |
| Pulselength | 6 ms | 3 ms |
| Pulse Drive Power | 26 kW | 85 kW |
| Filament Voltage | 13 V | 13 V |
| Filament Current | 900 A | 900 A |
| G1 DC-Voltage | pulsed from -700V to -600V | |
| G1 Pulse Current | 0.0 A | 4.6 A |
| Anode DC Voltage | 24.5 kV | 24.5 kV |
| Anode Pulse Current | 45 A | 120 A |
| G2 DC Voltage | 1450 V | 1450 V |
| G2 Pulse Current | 0.0 A | 4.0 A |
| Efficiency | 44 % | 78 % |

## 4 OPERATING PERFORMANCE

Six 200 kW and three 2 MW amplifiers were installed and brought to satisfying operation from January to August 1999. The implementation of a RF-feedback-system was not necessary because of the redesigned controls[3]. A hard to find malfunction of the control occurred at the IH1-amplifier: After some warm-up time a higher order resonance of the cavity drifted exactly to the $9^{th}$ harmonic, which influenced the reference values. A low pass filter in both reference lines solved this problem.

## 5 OUTLOOK

Some work will have to be done not to fire the crowbar if a cavity sparks which still leads to an averaged breakdown of a few minutes per month. We also intend to test the socket-compatible smaller RS2042SK of the old Wideroe amplifiers in the rebuilt prototype amplifier with additional vacuum capacitors. We still have four of those tubes and they are powerful enough to feed the RFQ.